\documentclass[twocolumn]{aastex631}
\usepackage{floatrow}
\usepackage{multirow}
\usepackage{bm}
\usepackage{xcolor}
\usepackage{array}
\newcommand{\change}[1]{{#1}}
\newfloatcommand{capbtabbox}{table}[][\FBwidth]
\begin{document}
\title{Tidal reconstruction of neutron star mergers from their late inspiral}
\author[0000-0003-2172-8589]{Souradeep Pal}
\affiliation{Indian Institute of Science Education And Research Kolkata \\
Mohanpur, Nadia - 741 246 \\
West Bengal, India}
\author[0000-0002-6814-7792]{K Rajesh Nayak}
\affiliation{Indian Institute of Science Education And Research Kolkata \\
Mohanpur, Nadia - 741 246 \\
West Bengal, India}
\begin{abstract}
We investigate the measurement correlation between the effective spin and the effective tidal deformability in gravitational wave signals from binary neutron star mergers. To efficiently measure the effective tidal deformability parameter, we exploit the fact that the tidal effects in a binary system are prominent when the components are closer during the late-inspiral. Thus, we indicate to a computationally efficient strategy of extracting the tidal information compressed within seconds before the merger. We report our observations for \texttt{GW170817} and explore the suitability of our approach in the upcoming observation scenarios of the current and the future ground-based observatories. Fast and accurate measurements of the tidal deformability parameters can be used to inform astronomers in prioritizing the electromagnetic follow-up efforts for such sources.
\end{abstract}
\section{Introduction} \label{sec:intro}
Even decades after the first observational evidence for the existence of  neutron stars, the pursuit for a consistent model for their equation of state continues in the broad astrophysics community. The equation of state (EoS), which describes how the matter inside the stars resists collapse against their own gravitational pull, can be constrained by their masses and the radii which are inferred through observations. Recent detections of the neutron star mergers by the current generation of gravitational wave observatories have opened up a new window of investigations.

The first three observing runs (O1-O3) of the advanced LIGO and the Advanced Virgo have reported two significant detections of binary neutron star (BNS) mergers \citep{abbott2017gw170817, Abbott_2020}. The joint detection and the association of gravitational waves (GW) and electromagnetic (EM) emissions for the event \texttt{GW170817} have put multi-messenger astronomy into the limelight \citep{Abbott_2017}. However, till now  \texttt{GW170817} is the only reported multi-messenger GW event observed by multiple observatories. A couple of neutron star - black hole (NSBH) systems have also been detected \citep{Abbott_2021_NSBH} but no EM counterparts were reported.

Accurate measurement of the tidal deformability in a neutron star caused by its companion object is singularly helpful in constraining the radius and the EoS of the star~\citep{abbott2018gw170817,PhysRevLett.121.091102,NeutronKaterina,Raithel_2018}. Additionally, these measurements can be used to probe the black-hole-nature of the components and provide valuable insights on the EM-properties, crucial for decision making by the astronomers during EM follow-ups.

Several techniques, with or without involving EM observations, have been used to measure the tidal effects primarily from the inspiral part of the GW signal \citep{PhysRevX.9.011001, dai2018parameter, Radice, PhysRevLett.121.091102}. While estimating the binary's parameters, Bayesian methods are generally robust, however, in the case of tidal deformability parameters, the inferences can be prone to variations from the choice of the prior model~\citep{PhysRevD.100.103023}. Nevertheless, full Bayesian analyses on fairly large populations of simulated BNSs are computationally challenging, given the long duration of the signals. Packages like \texttt{parallel\_bilby} offer fast schemes for inferring the binaries' properties, but can still take hours to process a single BNS signal even on a medium-sized computing cluster~\citep{pbilby}. Typically, a BNS signal spends hundreds of cycles in a detector's sensitive frequency band, however, the last several cycles to the merger would carry the best imprints of the tidal deformation effects. These interactions in a binary system are expected to be prominent when the components are in close proximity of each other before the merger. This is indicated in the literature~\citep{PhysRevD.85.123007,harry2018observing}. However, the standard approach is to match signals starting at lowest possible frequencies ($f\textsubscript{lower}\sim$ 20 Hz) by using the data containing the complete in-band inspiral~\citep{abbott2018gw170817,PhysRevLett.121.091102,PhysRevX.9.011001}.

In this work, we investigate the measurement correlation between the spins and the tidal deformability parameters in neutron star merger signals. We find that in the full frequency band of the detectors, spins can interfere with the accurate measurement of the tidal deformability parameters. Thus, we focus on the \textit{late-inspiral} part of the signals by starting the analyses at a higher frequency ($f\textsubscript{lower}\sim$ 150 Hz). The method uses the data available just seconds before the merger. Using simulations, we show that this approach can efficiently provide accurate estimates of the effective tidal parameter, $\lambda\textsubscript{eff}$ in the presence of the component spins. Here we consider the component spins that are aligned to the orbital angular momentum of the binary system. Note that $\lambda\textsubscript{eff}$ is better estimated than the individual tidal deformability parameters in a similar sense that the mass-weighted effective spin parameter, $\chi\textsubscript{eff}$ is more accurately measured than the individual aligned-spins. These two quantities represent effects that are not explicitly correleated and can be independently calculated using
\begin{equation}\label{eq:chieff}
\chi\textsubscript{eff}=\frac{m_{1}\chi_{1z}+m_{2}\chi_{2z}}{m_{1}+m_{2}}\,,
\end{equation}
\begin{equation}\label{eq:lameff}
\lambda\textsubscript{eff}=\frac{16}{13}\frac{m_{1}^{4}(m_{1}+12m_{2})\lambda_{1}+m_{2}^{4}(m_{2}+12m_{1})\lambda_{2}}{(m_{1}+m_{2})^{5}}
\end{equation}where $m_{i}$, $\chi_{i}$ and $\lambda_{i}$ represent the individual masses, the dimensionless spin parameters and the dimensionless tidal deformability parameters respectively.

There are several techniques to blend these effects together into waveform models. The classical approach using the post-Newtonian (PN) approximation is mostly valid in the low frequencies of the signal when the separation of the components is large~\footnote{\texttt{TaylorF2} terminates at the frequency corresponding to the innermost stable circular orbit~\citep{PhysRevX.9.011001}. Non-spinning tidal corrections enter the PN expansion at 5PN order.}. With the effective-one-body (EOB) formalism, it is possible to include the tidal corrections till the initial contact of the two bodies~\citep{nagar2018time}. In general, the coalescence of a neutron star merger becomes increasingly difficult to model as the components reach the merger, a regime where the numerical relativity (NR) informed approaches are expected to be more accurate~\citep{PhysRevD.81.084016,PhysRevD.99.024029,colleoni2023imrphenomxp_nrtidalv2}. It is also possible to incorporate the NR tidal corrections to precessing spin waveform models. However, parameter estimations using these waveform models have inferred only nominal spins for the GW events reported so far~\citep{PhysRevX.9.011001,dai2018parameter,PhysRevLett.121.091102,abbott2018gw170817,Abbott_2020}.

This work is organized as follows. In Section \ref{sec:method}, we discuss an overview of the GW searches for BNS systems involving spin as well as the tidal effects. We begin by exploring the benefits of including the tidal deformability parameters in the detection process. We investigate the deviations in the detection statistic for a set of simulated signals. We address the possibility of extracting the effective tidal deformability parameter at the detection stage from these simulations. Finally, we report the observed rates of background events with the inclusion of the tidal parameters using the detector data from the third Observing Run of LIGO-Virgo-KAGRA (LVK). We then discuss a novel method to efficiently decouple the effective spin and the effective tidal deformability measurements based on the frequency range of the analyses. The approach is extended to recovering the effective tidal deformability on a restricted frequency range. In Section \ref{sec:obs}, we report our observations for \texttt{GW170817} with the new approach and describe the effect of noise on these measurements. We also discuss a prospective strategy for such analyses in the upcoming observing scenarios. We make a brief remark on the computational cost of the analyses with the new approach.
\section{Method}\label{sec:method}
\setlength\extrarowheight{0.8pt}
\begin{flushleft}
\begin{table*}[t]
\begin{raggedright}
\begin{tabular}{>{\centering}m{4cm}>{\centering}m{1.8cm}>{\centering}m{1.8cm}>{\centering}m{1.8cm}>{\centering}m{0.1cm}>{\centering}m{1.8cm}>{\centering}m{1.8cm}>{\centering}m{1.8cm}}
\hline
\multirow{2}{2cm}{Search type} & \multicolumn{3}{c}{Injection space} &  & \multicolumn{3}{c}{Recovery space}\tabularnewline
\cline{2-4} \cline{3-4} \cline{4-4} \cline{6-8} \cline{7-8} \cline{8-8}
 & $m_{1},m_{2}(M_{\odot})$ & $\chi_{1},\chi_{2}$ & $\lambda_{1},\lambda_{2}$ &  & $m_{1},m_{2}(M_{\odot})$ & $\chi_{1z},\chi_{2z}$ & $\lambda_{1},\lambda_{2}$\tabularnewline
\hline
\hline
Small aligned-spins & $\mathcal{{U}}[1,3]$ & $\mathcal{{N}}[0,0.5]$ & $\mathcal{{U}}[0,4000]$ &  & $\mathcal{{U}}[1,3]$ & $\mathcal{{U}}[-0.05,0.05]$ & -\tabularnewline
Tidal only & $\mathcal{{U}}[1,3]$ & $\mathcal{{N}}[0,0.5]$ & $\mathcal{{U}}[0,4000]$ &  & $\mathcal{{U}}[1,3]$ & - & $\mathcal{{U}}[0,4000]$\tabularnewline
Large aligned-spins & $\mathcal{{U}}[1,3]$ & $\mathcal{{N}}[0,0.5]$ & $\mathcal{{U}}[0,4000]$ &  & $\mathcal{{U}}[1,3]$ & $\mathcal{{U}}[-0.95,0.95]$ & -\tabularnewline
Large aligned-spins + tidal & $\mathcal{{U}}[1,3]$ & $\mathcal{{N}}[0,0.5]$ & $\mathcal{{U}}[0,4000]$ &  & $\mathcal{{U}}[1,3]$ & $\mathcal{{U}}[-0.95,0.95]$ & $\mathcal{{U}}[0,4000]$\tabularnewline
\hline
\end{tabular}
\par\end{raggedright}
\begin{raggedright}
\caption{Intrinsic parameter spaces for the injections and the recovery by various simulated searches. Here $\mathcal{{U}}$ and $\mathcal{{N}}$ represent uniform and normal distributions. The extrinsic parameters for the injections are chosen from uniform random distributions in their respective domains. The distributions indicated for the recovery represent the initial distribution of the template points (also called \textit{particles}) in the PSO-based searches. The recovery of the detection statistic by the searches at a given template-sampling as described in the text is shown in Fig.~\ref{fig:snr_accuracy}.\label{tab:search_ranges}\vspace{0.5cm}}
\par\end{raggedright}
\end{table*}
\par\end{flushleft}
\vspace{-0.5cm}
To conduct searches for the candidate GW events and to perform parameter estimation on them, one essentially utilizes the inner product (or match) between the data and a model waveform~\citep{PhysRevD.44.3819,sathyaprakash2009physics}. The inner product between two vectors $a$ and $b$ is given by
\begin{equation}\label{eq:inner}
\langle{a|b}\rangle=4\Re{\int_{f\textsubscript{lower}}^{f\textsubscript{upper}}\frac{\tilde{a}(f)\tilde{b}^{*}(f)}{S\textsubscript{n}(f)}}df\,.
\end{equation}
Here $S\textsubscript{n}(f)$ represents the power spectral density (PSD) of the data which downweights the contribution of noise in the desired frequency domain $[f\textsubscript{lower}, f\textsubscript{upper}]$. A general problem is to find the true set of source parameters that maximizes the inner product. Here, we use particle swarm optimization (PSO) for obtaining a set of optimized source parameters by maximizing the inner product. PSO is a stochastic algorithm that iteratively optimizes a given \textit{fitness function} in an arbitrary dimensional parameter space~\citep{PhysRevD.81.063002,srivastava2018toward,PhysRevD.101.082001,pal2023swarm}. So the fitness function here is a quantity derived from the inner product, typically the signal-to-noise ratio (SNR) or the log-likelihood ratio (LLR) as described in detail in the following subsections.
\begin{figure*}[t]
\begin{floatrow}
\ffigbox{\includegraphics[scale=0.7]{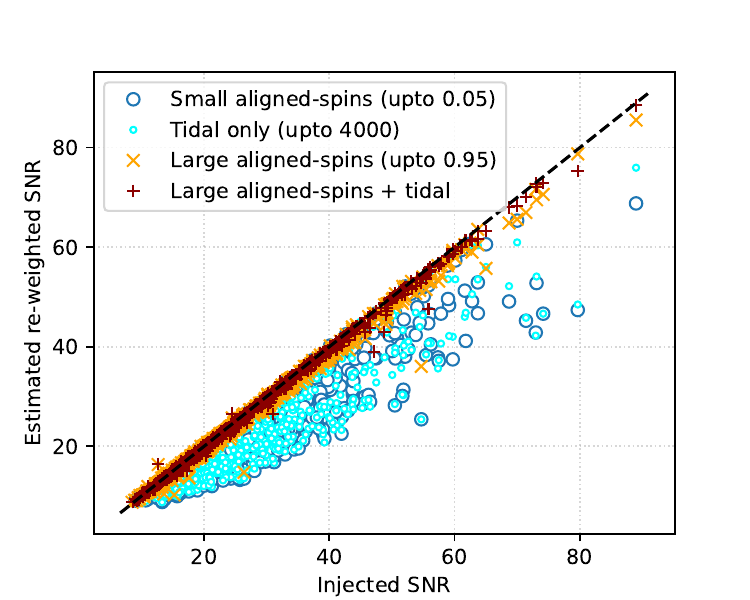}}
{\caption{Recovery of the detection statistic by various simulated searches at a fixed sampling. Each search runs on the common injection set of BNS signals simulated using \texttt{IMRPhenomXP\_NRTidalv2}. For simplicity, the searches are carried out with a 2 detector (HL) network. A time-coincidence test is used to arrive at the coincident detection statistic- the reweighted SNR for the coincident triggers which is computed using the standard $\chi\textsuperscript{2}$-veto~\citep{allen2005chi} as plotted on the y-axis. Data points lying close to the black dashed line represent optimal recovery of the injections.}
\label{fig:snr_accuracy}}
\vspace*{0.5cm}\ffigbox{\includegraphics[scale=0.7]{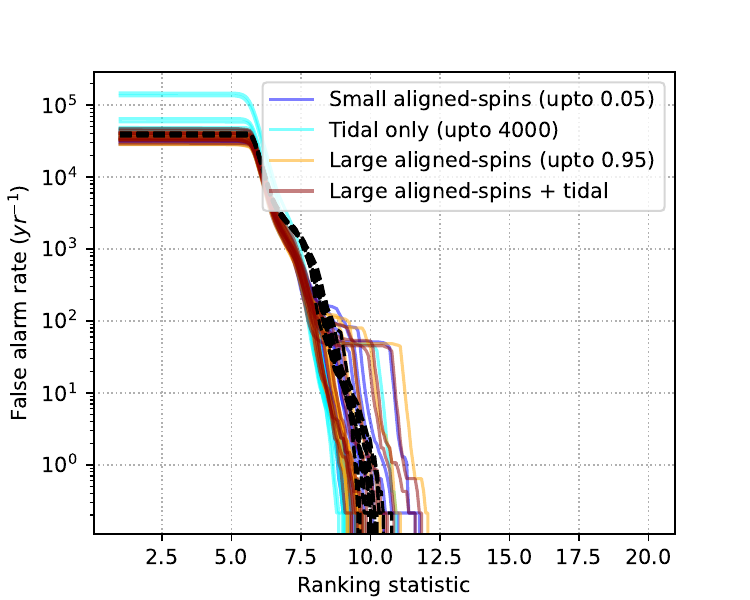}}
{\caption{Rates of background events compared for various searches obtained by running them on some 25 arbitrary HL coincident datasets from O3b. The background events are created by artificially time-sliding triggers from two or more detectors generated by the searches. The time slide duration is chosen to be larger than the maximum time taken by a GW signal to travel across detector, making the resulting events non-astrophysical. Searches identify astrophysical events by comparing their detection statistic against such background events from their local neighbourhood in time. The black dashed lines are for the searches on simulated Gaussian noise.}
\label{fig:roc_accuracy}}
\end{floatrow}
\end{figure*}
\subsection{Searches with the tidal parameters}\label{sec:search}
Gravitational wave searches constitute the first steps of astrophysical data processing whenever the analyzable data arrive from the observatories after calibration. The broad goal of the searches is to essentially identify \textit{events} of astrophyical origin from the strain timeseries data to enable detailed follow-up investigations on them. Our primary aim here is to investigate how the search process can possibly benefit from the inclusion of the tidal parameters and whether any tidal information can be extracted at the detection stage. Below we briefly discuss the analysis framework of matched-filter searches.

The candidate events need to be consistent with several astrophysical assumptions such as that the \textit{triggers} at different detectors caused by astrophysical events cannot be separated in time by more than the light's travel time between the detectors. Triggers are computed from the matched-filter SNR timeseries, $\rho(t)$, realizing Eq. \ref{eq:inner} with the strain data $d$ and a template waveform $h$
\begin{equation}\label{eq:mfo}
\rho^{2}(t;{\bm{\theta}})=4\Re{\int_{f\textsubscript{lower}}^{f\textsubscript{upper}}\frac{\tilde{d}(f)\tilde{h}^{*}(f;{\bm{\theta}})}{S\textsubscript{n}(f)}}e\textsuperscript{2{$\pi$}ift}df\,.
\end{equation}
The data from the observatories are cross-correlated with template waveforms that include various physical parameters ($\bm{\theta}$) describing the binaries. The number of the such parameters is often regarded as the search space dimensionality. Specific search configurations are adequately described in the literature~\citep{usman2016pycbc,adams2016low,messick2017analysis,PhysRevD.100.023011,chu2022spiir}. Searches for binary neutron stars in the Advanced LIGO - Advanced Virgo era have generally included small aligned-spins in addition to the component masses for the templates~\citep{Aubin_2021,ewing2024performance,Canton_2021}. A search for highly spinning light binaries was conducted recently but no additional significant event candidate was found~\citep{Wang_2024}. The first template bank search for neutron star mergers with the tidal parameters was conducted by \citep{chia2023pursuit}, however, their templates did not include any spin and also resulted into no additional significant detection. Instead the search reported even smaller SNRs for some of the events already detected by aligned-spin searches. Here, we test the detection algorithm based on PSO to include the tidal parameters in the presence of large aligned-spins. However, here we do not explicitly conduct a search targeting for real events over the archived LVK data and conclude based on simulations only.

We inject about 1000 simulated BNS signals starting at 20 Hz into Gaussian data segments with O4 model noise PSDs. These injections carry precessing spins with each spin component having a normal distribution around zero with the spin magnitudes restricted upto 0.95. The component masses and the tidal parameters are randomly chosen from $[1M_{\odot},3M_{\odot}]$ and $[0, 4000]$ respectively having a uniform distribution. Note that the population of the injections is not astrophysically informed, the primary goal here is to explore the capabilities of various searches. The signals are recovered with different combinations and/or ranges of template parameters using the PSO-based search algorithm. We use the \texttt{TaylorF2} model to generate templates choosing from a subset of the following parameters
\begin{equation}\label{eq:params}
\bm{\theta} =\{m_{1},m_{2},\chi_{1z},\chi_{2z},\lambda_{1},\lambda_{2}\}\,.
\end{equation}
Any combination of the search parameters includes at least the individual aligned-spins or the tidal deformability parameters in addition to the component masses. The largest search dimensionality explored is 6 which includes both the aligned-spin and the tidal parameters. The spin ranges are also allowed to vary for the aligned-spins. These are also summarized in Table.~\ref{tab:search_ranges}.

\begin{figure*}[t]
\begin{centering}
\includegraphics[scale=0.7]{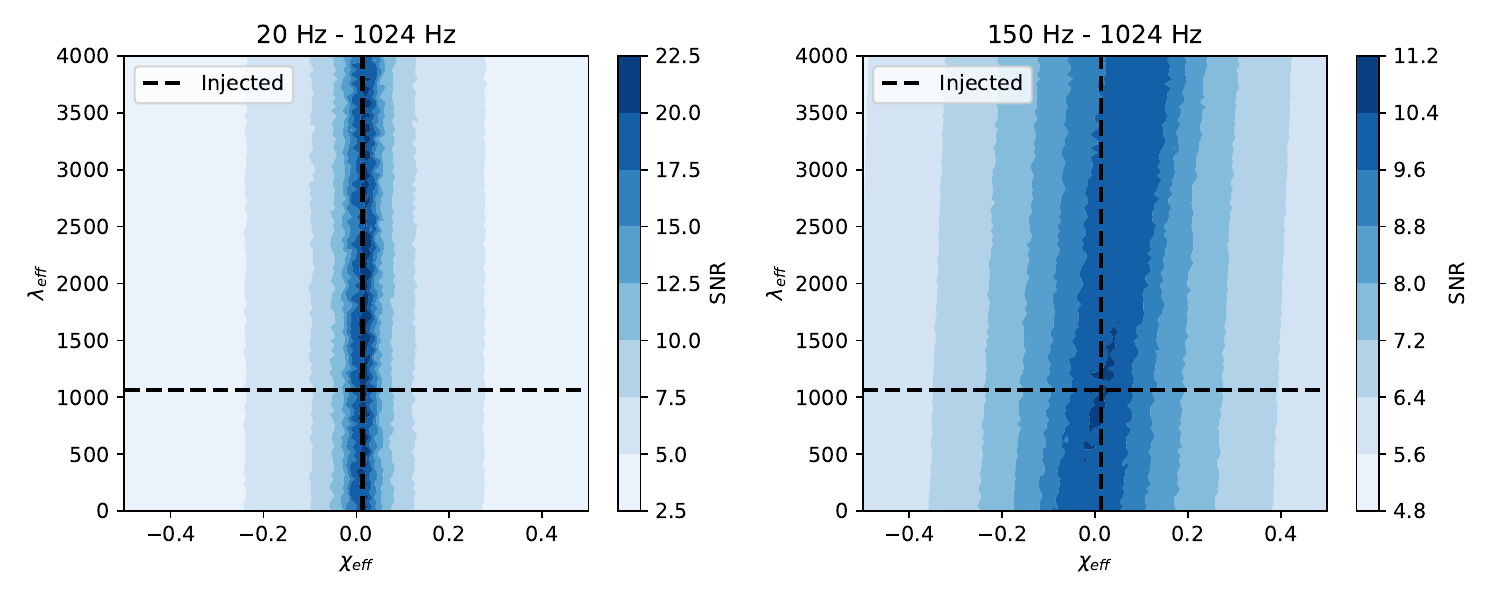}
\par\end{centering}
\caption{Tidal information as explicit for a typical BNS signal in the presence of aligned-spins. This is illustrated using SNR as the fitness function when the masses are fixed at the injected values. At $f\textsubscript{lower}\sim$ 20 Hz, while the fitness function peaks at the injected $\chi\textsubscript{eff}$, there is barely any information about the injected $\lambda\textsubscript{eff}$. At a higher $f\textsubscript{lower}\sim$ 150 Hz, the fitness function apparently becomes informed about the injected $\lambda\textsubscript{eff}$. Note that the central contour is slightly tilted in the right plot which almost aligns itself with the injected $\lambda\textsubscript{eff}$ as we further increase the $f\textsubscript{lower}$ as explored in Fig.~\ref{fig:lamvschi}. However, also note the broadening of the regions as the signal strength reduces with the signal bandwidth. A detailed description for obtaining these plots and the choice of the higher $f\textsubscript{lower}$ are presented in the subsequent subsection.\vspace{0.5cm}}
\label{fig:noinfo}
\end{figure*}
The PSO algorithm iteratively searches for an \textit{optimal} template point while searching in a given parameter space. The algorithm finally provides the corresponding detection statistic along with a set of the point estimates of the source parameters in a frequentist way. These are described in~\citep{pal2023swarm}. Concisely, the following two factors independently affect the recovery of the optimal statistic- (a) the waveform model, which ensures that the template is equipped with the physics to adequately model the incoming signal and (b) the template-sampling, which ensures that a sufficient matched-filter computations are performed throughout the search space in order to locate the peak value of the match function. For computational efficiency, a faster waveform model is preferred. On the other hand, the template-sampling is preset in a way that just recovers the signals at their injected \textit{statistic} values as shown in Fig. \ref{fig:snr_accuracy}. Note that the injected statistic value is obtained at the full injected source parameters, thus ideally, a search can achieve at most this value. To sufficiently sample any of the above parameter spaces, around 25500 template points are computed using the PSO algorithm. The template points are stochastically placed in the search parameter space on-the-fly during the search for any given injected signal. Since the search with large aligned-spins and tides recover the optimal detection statistic at this sampling (maroon pluses), any sub-optimal recovery by a restricted search cannot be due to undersampling~\footnote{However, a smaller sampling may be sufficient for a restricted parameter space.}. Thus, a smaller recovered statistic if any, reflects the fact that the restricted searches do not have access to the full search parameter space.

So these many points in the search parameter space are used to generate the template waveforms which are then used in the matched-filtering of the data from each detector. The resulting triggers from the individual detectors are processed with several consistency tests to obtain the \textit{coincident} events. The events are assigned a detection statistic from their reweighted SNRs and a false alarm rate is calculated based on time-slides~\citep{usman2016pycbc}. The methods adapted for the PSO algorithm are described in~\citep{pal2023swarm}.

\begin{figure}[b]
\begin{centering}
\includegraphics[scale=0.6]{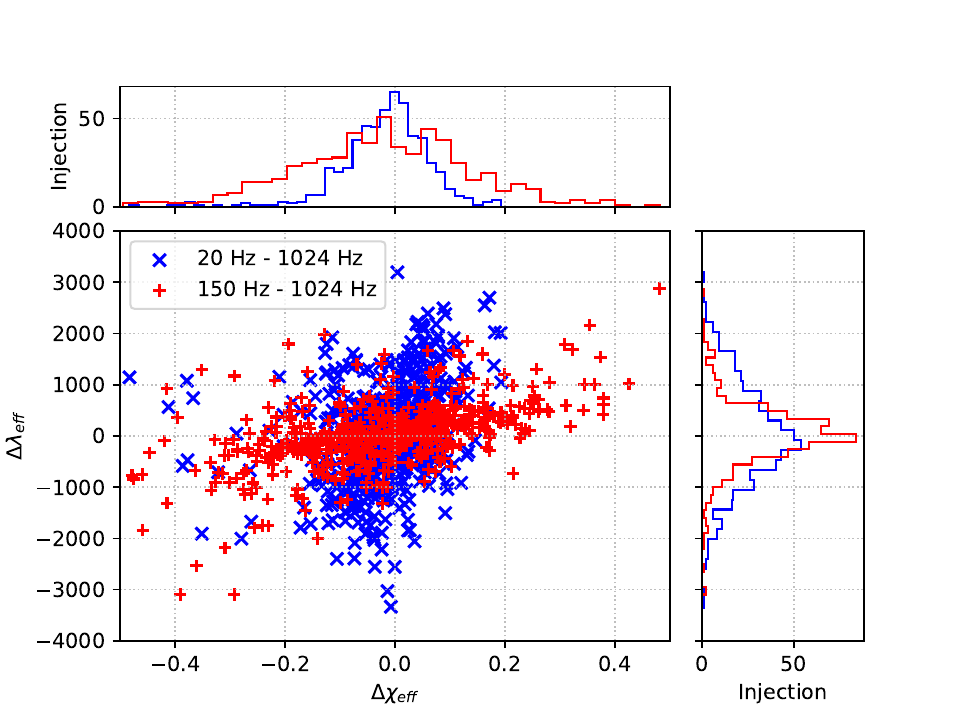}
\par\end{centering}
\caption{The correlation between the errors in the point estimates for $\chi\textsubscript{eff}$ and $\lambda\textsubscript{eff}$ obtained from the six-dimensional PSO search including spins and tides at two different $f\textsubscript{lower}$'s. At the same detection criterion, a smaller fraction of the total injected signals are recovered while using a higher $f\textsubscript{lower} \sim$ 150 Hz as expected. However, the point estimates of $\lambda\textsubscript{eff}$ for the commonly found injections are generally more accurate than those for $f\textsubscript{lower} \sim$ 20 Hz.}
\label{fig:PSO_lamvschi}
\end{figure}
\begin{figure*}[t]
\begin{centering}
\includegraphics[scale=0.58]{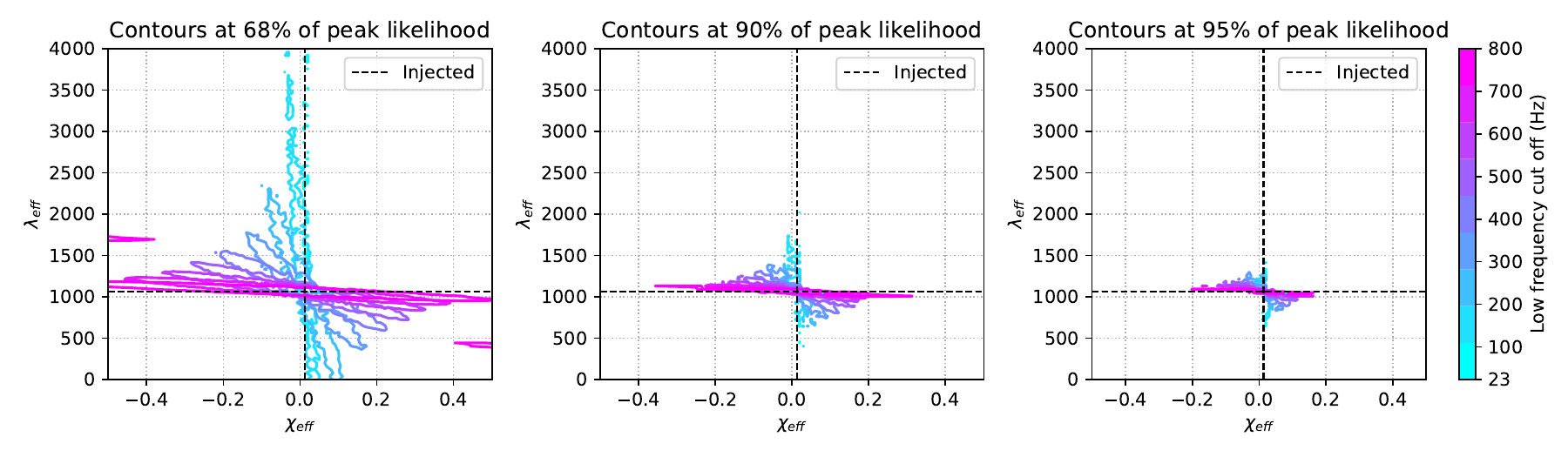}
\par\end{centering}
\caption{Contours containing the top likelihood points in a grid of $\lambda\textsubscript{eff}$ vs $\chi\textsubscript{eff}$ as a function of the low frequency cut-off for the template waveforms. A contour at x\% of the peak likelihood encloses all the points in the grid that have a likelihood value greater than or equal to x\% of the peak likelihood value in the grid. The plots at three different contour levels are shown. The component masses are fixed to the injected values of a typical BNS source ($1.4M_{\odot}, 1.4M_{\odot}$) located at 200 Mpc.\vspace*{0.5cm}}
\label{fig:lamvschi}
\end{figure*}
\begin{figure}[b]
\begin{centering}
\includegraphics[scale=0.6]{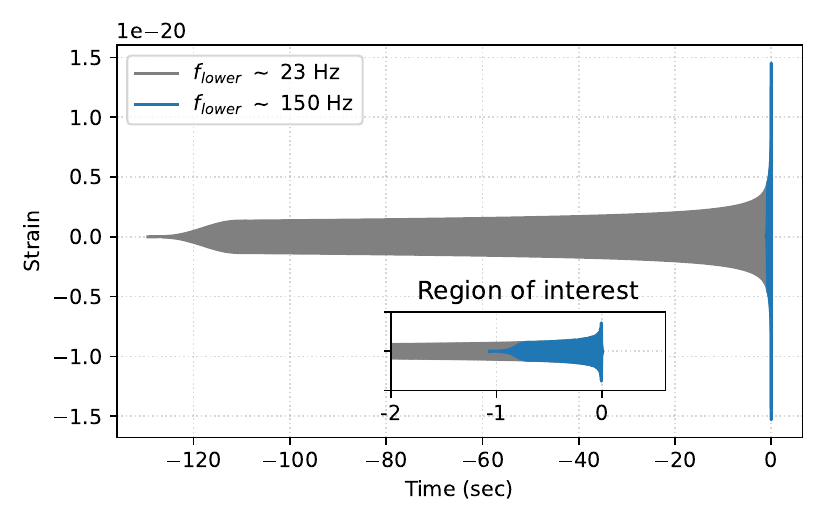}
\par\end{centering}
\caption{Visualization of the template signals at ($1.4M_{\odot}, 1.4M_{\odot}$) in the time domain used to extract the spin and the tidal effects during the detection stage. Detecting signals at a smaller $f\textsubscript{lower}$ (in grey) is desirable and effective for capturing the spin effects, while a larger $f\textsubscript{lower}$ (in blue) can be used for the portion of the signal when the tidal interactions are prominent. The later analysis requires significantly smaller duration of strain data close to the merger and is thus fast and reliable for the tidal deformation effects.\vspace*{0.5cm}}
\label{fig:wav}
\end{figure}
\change{When the search parameter space is expanded, the resulting  background event rates can potentially change. In the case of template bank searches, the collection of triggers resulting from a larger template bank is also larger, increasing the background rates. Such effects have been explored while including effects such as spin-precession, for example in Gaussian noise~\citep{PhysRevD.94.024012,PhysRevD.97.023004,PhysRevD.106.103035}. On the other hand, extended parts of a larger parameter space can also respond abruptly to the occasional glitchy data from the detectors. Usually, the signal-based power $\chi\textsuperscript{2}$-test is capable of effectively downranking any such triggers from long duration templates for the lighter binaries~\citep{allen2005chi}. The processes used to obtain the triggers from the matched-filter outputs can further decide the significance assigned to the candidate events. In the PSO-based searches, the significance of a candidate is calculated by buffering the triggers resulting from the optimized templates of the local data stretches. Since only a single (the optimized) template is used, the background depends on the convergence point in the search parameter space and is nearly independent of the number of template points used to obtain the convergence. This is approximately valid at a sufficient template-sampling. Thus, the background estimation is largely independent of the template-sampling used. This method is described and validated in detail in~\citep{pal2023swarm}. Here we skip reporting the background estimation procedure and proceed with the observations.}

\change{To demonstrate how the search configurations described earlier respond to instrumental noise and generate the background statistic, we exercise the PSO-based searches on real datasets from O3b that do not contain any reported GW signal or known hardware injection~\citep{RICHABBOTT2021100658}. These datasets are 4096 seconds in duration and the searches use an identical setup as described earlier. We observe that each of the search configurations respond to the noise events from the detector data in a similar fashion, resulting into similar rates of non-astrophysical background events, as summarized in Fig. \ref{fig:roc_accuracy}. We note that the background event rates are fairly consistent with that of the Gaussian noise case and obtain no excess of triggers (rate or loudness) in these datasets.} A summary of the simulation results is the following.
\begin{enumerate}
\item[\textbullet]Between spin and tide, the former plays the dominant role in obtaining the optimal detection statistic. The tidal parameters generally have a small contribution to the build-up of the detection statistic.
\item[\textbullet]We find that the templates used in searches should have at least large aligned-spins to optimally recover sources that may have generic spins. A search with templates having tidal parameters but no spin can lose a significant fraction of the retrievable detection statistic.
\end{enumerate}
Thus we conclude that the search parameter space \textit{should} include the tidal parameters in the presence of large generic aligned-spins to approach the optimal sensitivity through these parameters toward the observable binary neutron stars. But in the present observing scenario, the advantage of including the tidal deformability parameters (upto $\sim$4000 as supported by most waveform models) may not be significant in most cases. With the PSO-based approach, the computational cost of implementing such a search is reasonably trivial. However, the information content for the tidal deformability effects is \textit{screened} by the effective spin as shown in Fig.~\ref{fig:noinfo}. Consequently, the search with aligned-spins and the tidal parameters reports relatively more accurate point estimates for the effective spin than for the effective tide in the full frequency range of the detectors as shown in the blue crosses in Fig.~\ref{fig:PSO_lamvschi}. This is taken up further in the next subsection where we discuss the origin of the accuracy of the red pluses in detail.

\begin{figure*}[t]
\begin{centering}
\includegraphics[scale=0.7]{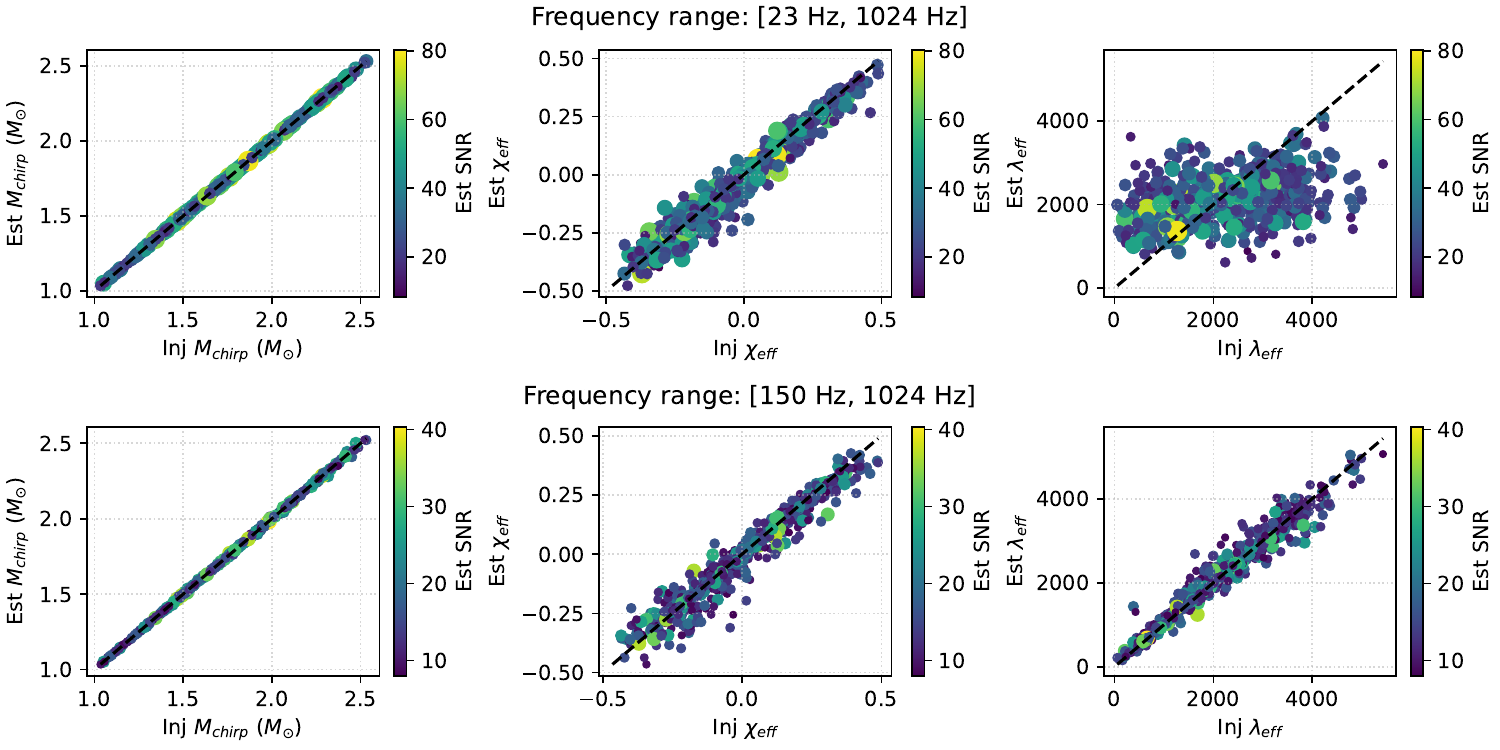}
\par\end{centering}
\caption{Summary of the point estimates of the intrinsic parameters of multiple simulated BNS sources in zero noise recovered by the PSO search. The search parameters are ($m_{1}, m_{2}, \chi_{1z}, \chi_{2z}, \lambda_{1}, \lambda_{2}$) whereas other remaining parameters were fixed at their injected values. The injection and recovery are carried out using the \texttt{IMRPhenomD\textunderscore{NRTidal}} waveform model. The top and the bottom panels are for the two frequency ranges [23 Hz - 1024 Hz] and [150 Hz - 1024 Hz] analyzed with the configuration otherwise. From the rightmost plots, we note that the errors in the point estimates for $\lambda\textsubscript{eff}$ are smaller with the $f\textsubscript{lower}$ of 150 Hz than with the $f\textsubscript{lower}$ of 23 Hz.\vspace*{0.5cm}} \label{fig:PSO_likelihood}
\end{figure*}
\subsection{Estimation of the effective tidal parameter}\label{sec:estimate}
In this section, we first discuss the measurement correlation between the effective spin and the effective tide in detail through the likelihood function. The likelihood ($\mathcal{L}$) of a signal with the parameters $\bm{\theta}$ in the data $d$ assuming a Gaussian noise model is given as
\begin{equation}\label{eq:likelihood}
\mathcal{L}=e^{-\langle{d-h(\bm{\theta})|d-h(\bm{\theta})}\rangle/2}\,,
\end{equation}
where $\langle{a|b}\rangle$ represents the inner product of two vectors $a$ and $b$ defined in Eq. \ref{eq:inner}. The log-likelihood ratio is often used to quantify a competing signal hypothesis against the noise hypothesis which simplifies to
\begin{equation}\label{eq:loglr}
LLR=\langle{d|h(\bm{\theta})}\rangle-\frac{1}{2}\langle{h(\bm{\theta})|h(\bm{\theta})}\rangle\,.
\end{equation}
Information about the true signal parameters can be extracted by maximizing the likelihood function. In the Bayesian approach, it is used to generate the posterior probability distributions for a desirable set of the parameters. However, we do not use a Bayesian approach in this work to avoid the explicit use of priors. Instead our approach here is to maximize the likelihood itself. We use the PSO algorithm to obtain the maximum likelihood estimate (MLE). So given an initial distribution of particles, as long as the PSO algorithm recovers the maximum likelihood, the recovered MLE parameters are expected to be sane. We use \textit{error} to denote the difference between an estimated value and the true value of a quantity if it is known, e.g. in simulations. When the true value is not known, e.g. in a real scenario, we denote the diffidence by \textit{uncertainty}.
\subsubsection{Grid based study}\label{sec:grid}
While measuring, several of the source parameters can be degenerate with one another- the measurement of one can interfere with the measurement of another. Inherently, several combinations of the degenerate parameter values lead to almost the same likelihood values, confusing the numerical algorithms. To understand the behaviour of the likelihood function, we construct a grid of $\lambda\textsubscript{eff}$ versus $\chi\textsubscript{eff}$ and calculate the likelihoods at the grid points for a simulated BNS signal while keeping other signal parameters fixed at their injected values. The injected signal is for a BNS source with component masses ($1.4M_{\odot}, 1.4M_{\odot}$) at 200 Mpc generated using the \texttt{IMRPhenomD\textunderscore{NRTidal}} waveform model. The signal is projected onto a simulated HL detector network (O3). For the grid-sampling, we choose a moderate range of $\chi\textsubscript{i}$ upto $\pm0.5$ and of $\lambda\textsubscript{i}$ upto $4000$ to generate an equally spaced grid of points in the $\chi\textsubscript{eff}$-$\lambda\textsubscript{eff}$ plane. These points are obtained by arbitrarily varying the values of $\chi\textsubscript{i}$ and $\lambda\textsubscript{i}$ for the injected component masses $m\textsubscript{i}$ and are preserved throughout the simulation. Finally the LLR is computed at all the grid points and contours containing the top likelihood points are plotted. The likelihood computations are repeated by changing the value of $f\textsubscript{lower}$ as shown in Fig. \ref{fig:lamvschi}. The contour levels are varied across the subfigures. We observe that the contours at a given level contain a larger range of values for $\lambda\textsubscript{eff}$ than for $\chi\textsubscript{eff}$ when the waveforms are computed at the smaller $f\textsubscript{lower}$'s (blues). Thus the peaks of likelihood function are sharper near the injected $\chi\textsubscript{eff}$ and broader near the injected $\lambda\textsubscript{eff}$ in these cases. This trend keeps reversing on increasing the $f\textsubscript{lower}$ with the likelihood function becoming sharper toward $\lambda\textsubscript{eff}$ and the contours orient along the injected value of $\lambda\textsubscript{eff}$ (pinks). We find that this unique behaviour of the likelihood function is independent of the injected values of the spins and tides and is consistent at all signal strengths. However, note that the likelihood value diminishes with increasing $f\textsubscript{lower}$ due to the loss of the signal strength in the relevant frequency band. A depiction of the changing signal duration is given in Fig.~\ref{fig:wav}.
\subsubsection{PSO-based likelihood optimization}\label{sec:PSO_likelihood}
In the above simulation, we assumed the signal parameters other than the spins and the tides to be fixed at the injected values for computing the likelihood. Next we investigate by releasing the component masses as \textit{free} parameters. Computing the likelihood (or the LLR) function on the resulting instrinsic parameter space with a uniform grid is computationally expensive and not optimal. Hence, we use the PSO algorithm to search over the intrinsic parameter space with the LLR as the fitness function. However in the real cases, all the signal parameters need to be treated as free parameters and be searched over their possible domains. This is discussed further in the next section.

Multiple random injections are generated using the \texttt{IMRPhenomD\_NRTidal} waveform model and are recovered by the PSO algorithm in two selected frequency bands, namely [23 Hz - 1024 Hz] and [150 Hz - 1024 Hz] with the same waveform model. These are summarized in Fig. \ref{fig:PSO_likelihood}. The injection distribution of this simulation is similar to that mentioned in the previous section except that these are non-precessing injections but with uniform aligned-spin components. To isolate the effect of changing the $f\textsubscript{lower}$, these simulations are performed in zero noise. We observe that the point estimates for $\lambda\textsubscript{eff}$ with the $f\textsubscript{lower}$ of 150 Hz are more accurate than that with the $f\textsubscript{lower}$ of 23 Hz. We attribute the result to the underlying unique behaviour of the likelihood function in the spin-tide space as explored earlier.
\subsubsection{Choice of the low frequency cutoff}\label{sec:flower}
\begin{figure}[t]
\begin{centering}
\includegraphics[scale=0.6]{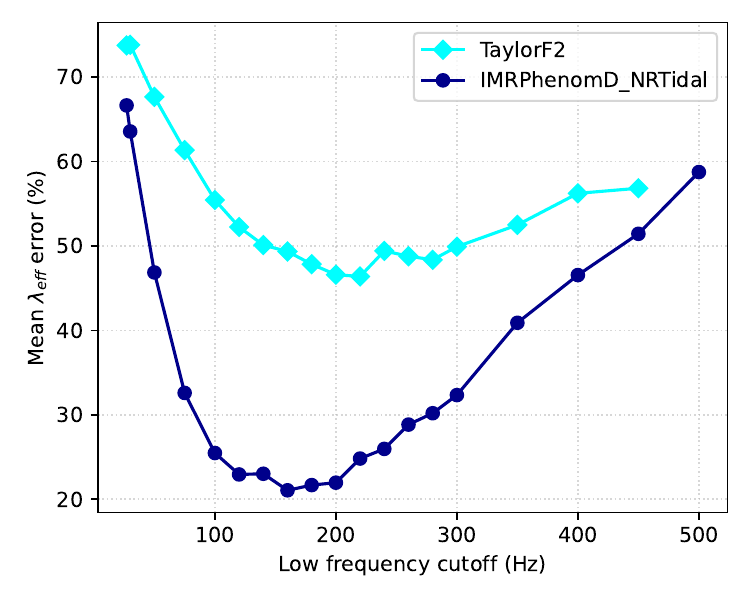}
\par\end{centering}
\caption{Errors in the point estimates of $\lambda\textsubscript{eff}$ averaged over 10k simulated sources plotted as a function of the $f\textsubscript{lower}$ in O3-like sensitivity.}
\label{fig:errors_vs_freq}
\end{figure}
An optimal choice of the $f\textsubscript{lower}$ is investigated through simulations as described below. We generate a set of 10k random BNS injections and recover them as described above. We use different values of $f\textsubscript{lower}$ starting at $\sim$20 Hz till $\sim$500 Hz for recovering the same set of injections. We plot the average error in $\lambda\textsubscript{eff}$ for the injection polpulation as a function of the $f\textsubscript{lower}$ as shown in Fig.~\ref{fig:errors_vs_freq}. There are two main competing factors here- the alignment and the broadening of the contours as indicated in Fig.~\ref{fig:lamvschi}. At small $f\textsubscript{lower}$'s (below 100 Hz), the measurement of $\chi\textsubscript{eff}$ overwhelms the measurement of $\lambda\textsubscript{eff}$ since the contours are aligned along the true $\chi\textsubscript{eff}$. At large $f\textsubscript{lower}$'s (beyond 300 Hz), the fall in the signal strength broadens the contours which inhibits any further improvement with a possibly higher $f\textsubscript{lower}$. Thus, we find that an intermediate value of $f\textsubscript{lower}$ is optimal which minimizes the error in $\lambda\textsubscript{eff}$ for the given set of simulated sources. Note that this optimum value for the $f\textsubscript{lower}$ is empirical and can slightly vary depending on the population chosen for the injections, the PSDs and other physical configurations. However, we assume an $f\textsubscript{lower}$ of 150 Hz from here onwards in this work. Here two different waveform models are tested- \texttt{TaylorF2} and \texttt{IMRPhenomD\textunderscore{NRTidal}}. Note that the later, being informed of the accurate NR-tidal corrections, is expected to give more reliable estimates, in particular with the new approach based on the late-inspiral.

\setlength\extrarowheight{0.5pt}
\begin{table*}[t]
\begin{tabular}{>{\centering}m{5.5cm}>{\centering}m{2.1cm}>{\centering}m{2.1cm}>{\centering}m{0.25cm}>{\centering}m{2.1cm}>{\centering}m{2.1cm}}
\hline
 & \multicolumn{2}{c}{Injection space} &  & \multicolumn{2}{c}{Recovery space}\tabularnewline
\cline{2-3} \cline{3-3} \cline{5-6} \cline{6-6}
Parameter & Minimum & Maximum &  & Minimum & Maximum\tabularnewline
\hline
\hline
Primary mass, $m_{1}$ $(M_{\odot})$ & 1 & 3 &  & 1 & 3\tabularnewline
Secondary mass, $m_{2}$ $(M_{\odot})$ & 1 & 3 &  & 1 & 3\tabularnewline
Primary z-spin, $\chi_{1z}$ & -0.5 & 0.5 &  & -0.5 & 0.5\tabularnewline
Secondary z-spin, $\chi_{2z}$ & -0.5 & 0.5 &  & -0.5 & 0.5\tabularnewline
Primary tidal deformability, $\lambda_{1}$ & 0 & 4000 &  & 0 & 4000\tabularnewline
Secondary tidal deformability, $\lambda_{2}$ & 0 & 4000 &  & 0 & 4000\tabularnewline
Coalescence time, $t_{c}$ (GPS seconds) & 1100000000 & 1500000000 &  & $t_{c}-0.01$ & $t_{c}+0.01$\tabularnewline
Distance, $D_{L}$ (Mpc) & 50 & 60 &  & $D_{L}-30$ & $D_{L}+30$\tabularnewline
Inclination, $\iota$ (rad) & 0 & $\pi$ &  & $\iota-0.1$ & $\iota+0.1$\tabularnewline
Right ascension, $\alpha$ (rad) & 0 & $2\pi$ &  & $\alpha-0.1$ & $\alpha+0.1$\tabularnewline
Declination, $\delta$ (rad) & $-\pi/2$ & $\pi/2$ &  & $\delta-0.05$ & $\delta+0.05$\tabularnewline
Coalescence phase, $\phi$ (rad) & 0 & $2\pi$ &  & $0$ & $2\pi$\tabularnewline
Polarization, $\psi$ (rad) & 0 & $2\pi$ &  & \multicolumn{2}{c}{Marginalized}\tabularnewline
\hline \caption{Injection and estimation ranges of the signal parameters used for likelihood optimization using the PSO algorithm. The intrinsic parameters are searched over their respective possible domains while the extrinsic parameters are searched over a small window around their true values or can be marginalized over. The results in this work are obtained with these settings unless stated otherwise.\vspace*{0.5cm}}
\label{tab:ranges}
\end{tabular}
\end{table*}
\begin{figure}[b]
\begin{centering}
\includegraphics[scale=0.61]{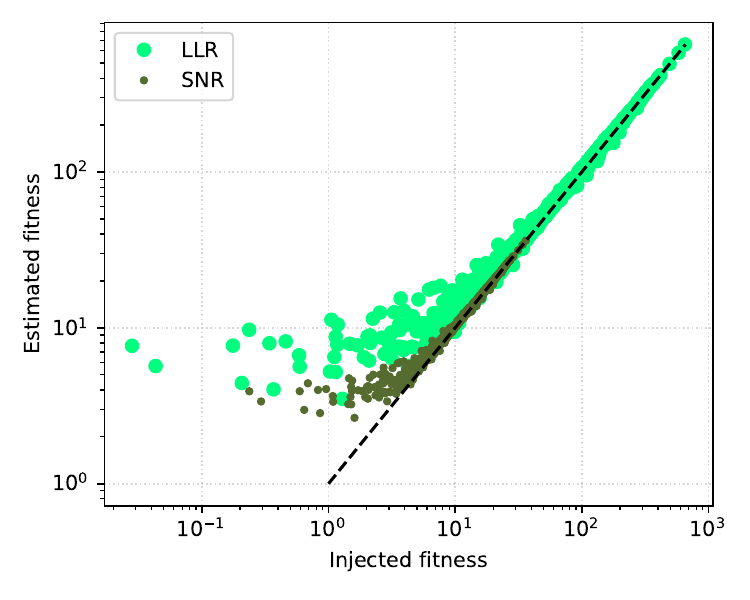}
\par\end{centering}
\caption{Recovery of multiple random injections with a single swarm exploring the intrinsic parameter space plus a small window around the extrinsic parameters as described in Table~\ref{tab:search_ranges}. We optimize the LLR as the fitness function but also show the SNR values corresponding to the obtained LLR values.}
\label{fig:sinfit}
\end{figure}
\section{Observations}\label{sec:obs}
\begin{figure}[b]
\begin{centering}
\includegraphics[scale=0.61]{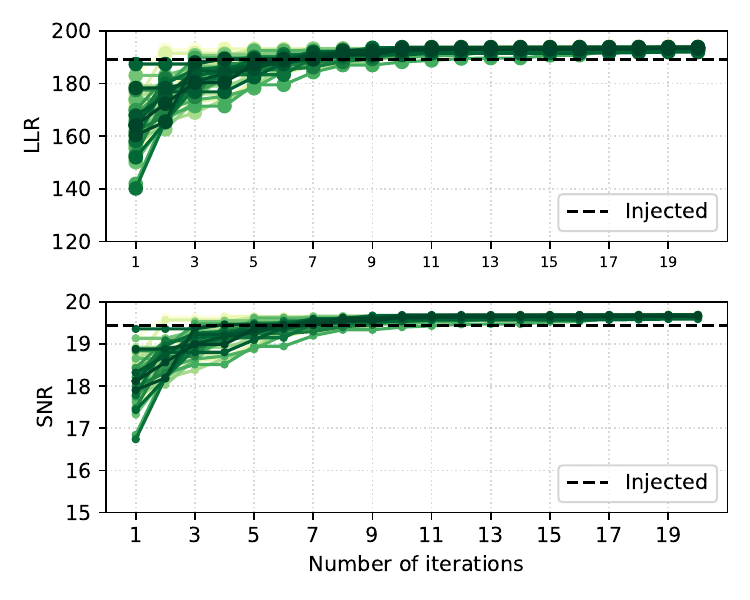}
\par\end{centering}
\caption{A GW170817-like injection recovered from simulated Gaussian noise using PSO. Each curve represents a swarm independently exploring the search parameter space as described in the text. The plot indicates that the algorithm picks up the optimal fitness value within 20 iterations while using $\sim$ 10000 particles per iteration. The injected fitness is computed at the injected signal parameters.}
\label{fig:mulfit}
\end{figure}
\begin{figure*}[t]
\begin{centering}
\includegraphics[scale=0.6]{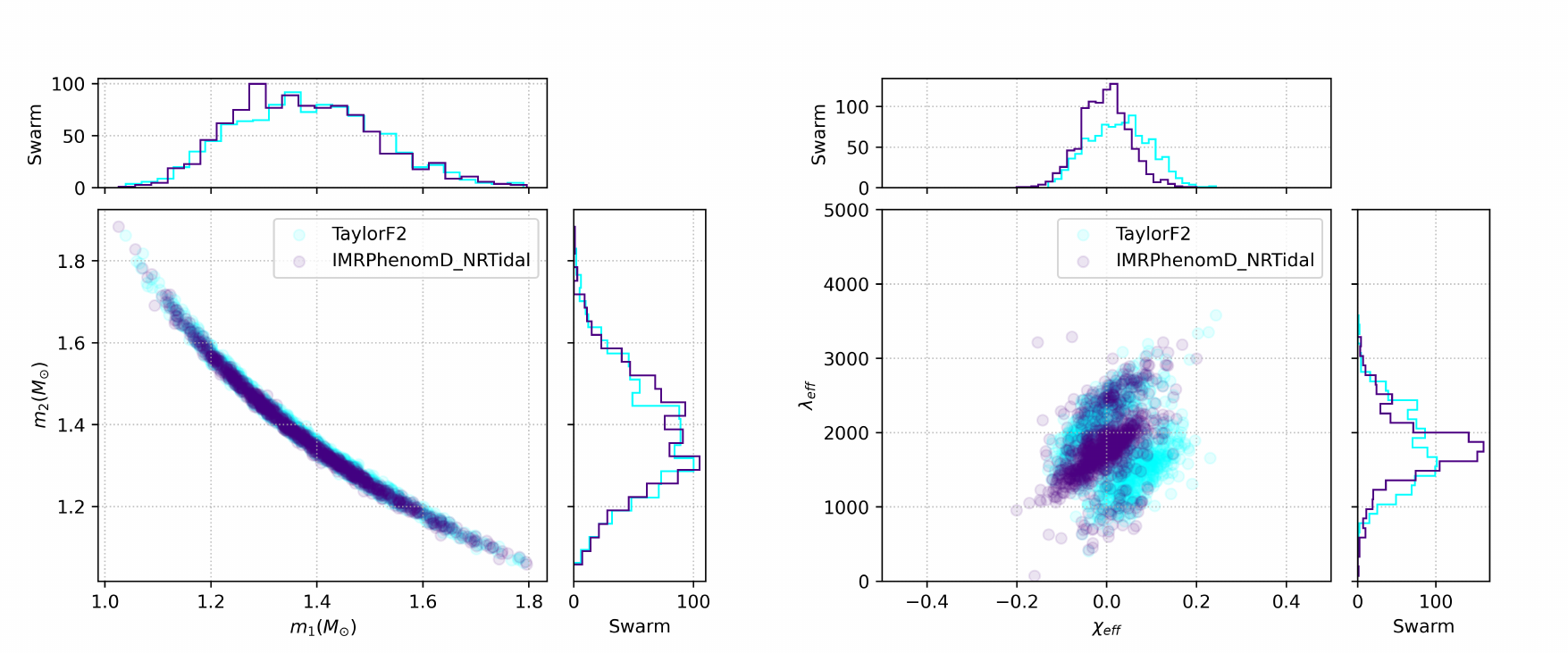}
\par\end{centering}
\caption{Parameters reconstructed from the optimal template points reported by the multiple swarms that are used to independently explore the \texttt{GW170817} event in the frequency range 150 Hz - 1024 Hz. Only the intrinsic parameters are shown. The extrinsic parameters like the sky location and the luminosity distance are sampled around the values reported from EM observations or are marginalized over as indicated in Table~\ref{tab:ranges}.\vspace*{0.5cm}}
\label{fig:GW170817_multi_swarm}
\end{figure*}
In the previous section, we explored a novel approach to obtain a point estimate of the tidal deformability through a restricted frequency range. In this section, we discuss the applicability of this approach in real observing scenarios.

Earlier, we fixed the extrinsic parameters to their injected values. In other words, we assumed that the extrinsic parameters to be perfectly known. In the most general case, all the signal parameters need to be treated as free parameters and be searched over their possible domains. However, for the current scope, we are interested in obtaining the intrinsic parameters accurately, specifically the tidal parameters. In practice, this can be tackled with a suitable marginalization scheme or by roughly treating some parameters as \textit{semi-free} parameters that are available from external observations, if known. For example, in the case of \texttt{GW170817}, one can combine the information from the EM observations and use the known sky position to sample over a smaller window for optimization. On the other hand, the coalescence time of the event can be sampled around the trigger time of the GW event which is a standard practice in GW parameter estimation. A summary of this scheme is given in Table.~\ref{tab:ranges}. Thus, we carry out another injection study by turning on Gaussian noise. To tackle the increased parameter space coverage by allowing small windows for the extrinsic parameters, we configure the PSO algorithm with around 10000 particles and 20 iterative steps. At this sampling, the efficacy of the algorithm in recovering the LLR as compared to the injected LLR for the random injections is shown in Fig. \ref{fig:sinfit}.

While the algorithm stochastically locates the maximum likelihood (or LLR) through a point estimate, the parameter values of the optimally matching template are prone to statistical variations from several expected origins. The optimal point, and hence the convergence point assuming the PSO algorithm performs ideally, can differ from the true source parameters depending on the random noise realization in which the signal is deposited. Additionally, the algorithm can itself report a slightly different point of  convergence from the optimal as a deviation from ideality. Nevertheless, the optimization can be considered acceptable, provided that the swarm reports the optimal likelihood (or LLR) value within a desired tolerance~\footnote{Alternately, one can always preset the template-sampling in order to tolerate only a small desirable loss in the recovered LLRs for a given set of injections.}.

To address the issue of the uncertainty of a point estimate, we devise a \textit{multi-swarm} strategy where the algorithm independently explores a given injection multiple times. The explorations are carried out at the same search configuraion but begining with a distinct seed~\footnote{The algorithm just chooses a distinct path, eventually converging on the contour of the optimal likelihood.}. Thus, every exploration is equivalent to any other in a frequentist sense. This is demonstrated for a \texttt{GW170817}-like injection in Fig.~\ref{fig:mulfit}. The recovered LLR (or the SNR) values from all the individual explorations agree to a reasonable extent at the end of the iterations. To infer the source parameters, the point estimate obtained from each exploration is used.

Note that with a higher $f\textsubscript{lower}$, the computational cost of each of the analysis becomes reasonably small mainly due to (a) the smaller duration of the data that is analyzed, $\sim4$ seconds before the merger and (b) the faster generation of template waveforms due to the smaller number of inspiral cycles computed. We find that these analyses take seconds (or minutes) on multi-core (or single-core) machines. Fortunately, the reduced computational cost keeps the multi-swarm approach computationally practicable.

\begin{figure*}[t]
\begin{centering}
\includegraphics[scale=0.75]{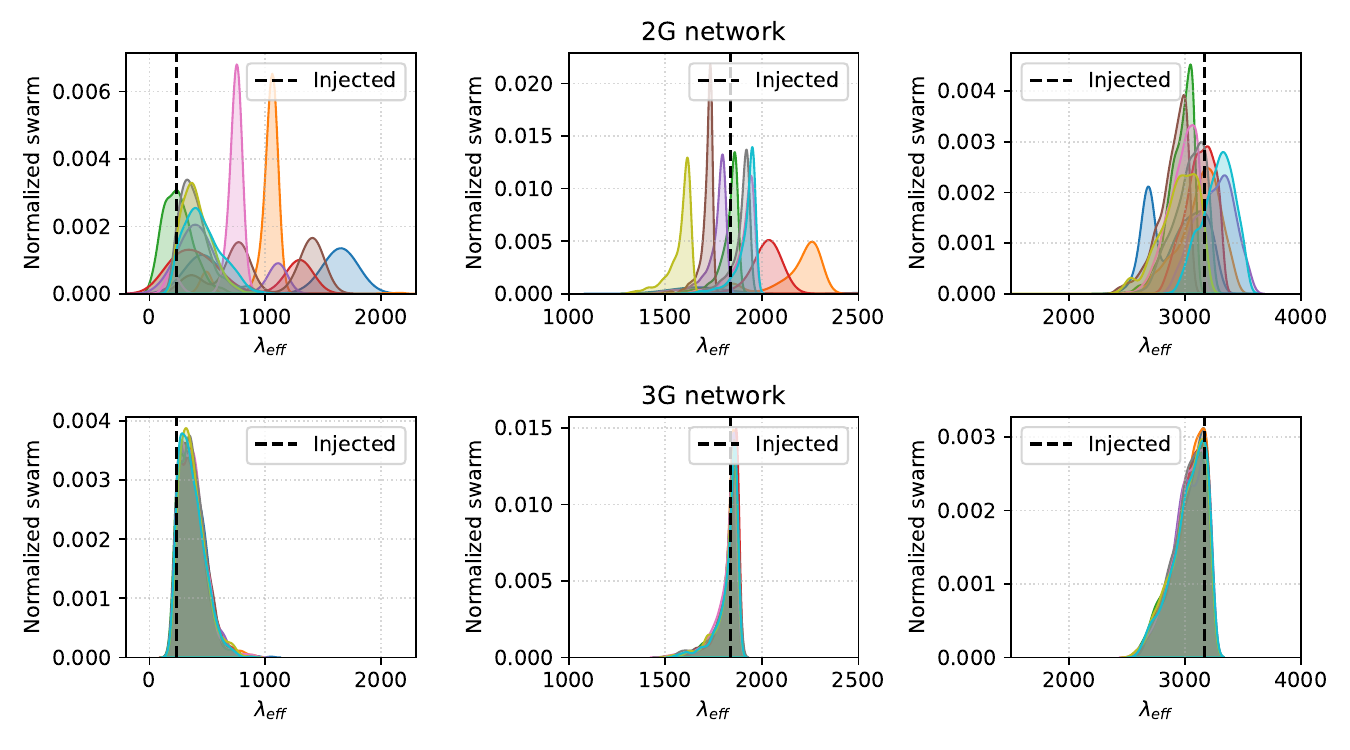}
\par\end{centering}
\caption{Effect of detector noise realization on the inference of $\lambda\textsubscript{eff}$ obtained with multi-swarm analyses in the [150 Hz, 1024 Hz] frequency band for a simulated \texttt{GW170817}-like source. We inject three types of $\lambda\textsubscript{eff}$ values for the same source otherwise- small (left), moderate (middle) and large (right), each into ten different Gaussian noise realizations. Two different observation scenarios are shown- (top) LIGO Hanford-LIGO Livingston network at O3-like sensitivities and (bottom) Cosmic Explorer-Einstein Telescope network at the projected sensitivities, having the in-band network SNRs of $\sim$15 and $\sim$285 respectively. These are recovered by the multi-swarm analysis and the inference are plotted each with a different color. To give the essence better, the histograms for the normalized swarm count are smoothened. With these uncertainty estimates, the inference is considered reliable when the histograms peak at the injected values. Note that the histograms consistently peak near the true value in the 3G case, so for the purpose of low-latency computations such as source classification, the peak value could be used until a full Bayesian estimate becomes available.\vspace*{0.5cm}}
\label{fig:noise_inference}
\end{figure*}
\begin{figure*}[t]
\begin{centering}
\includegraphics[scale=0.7]{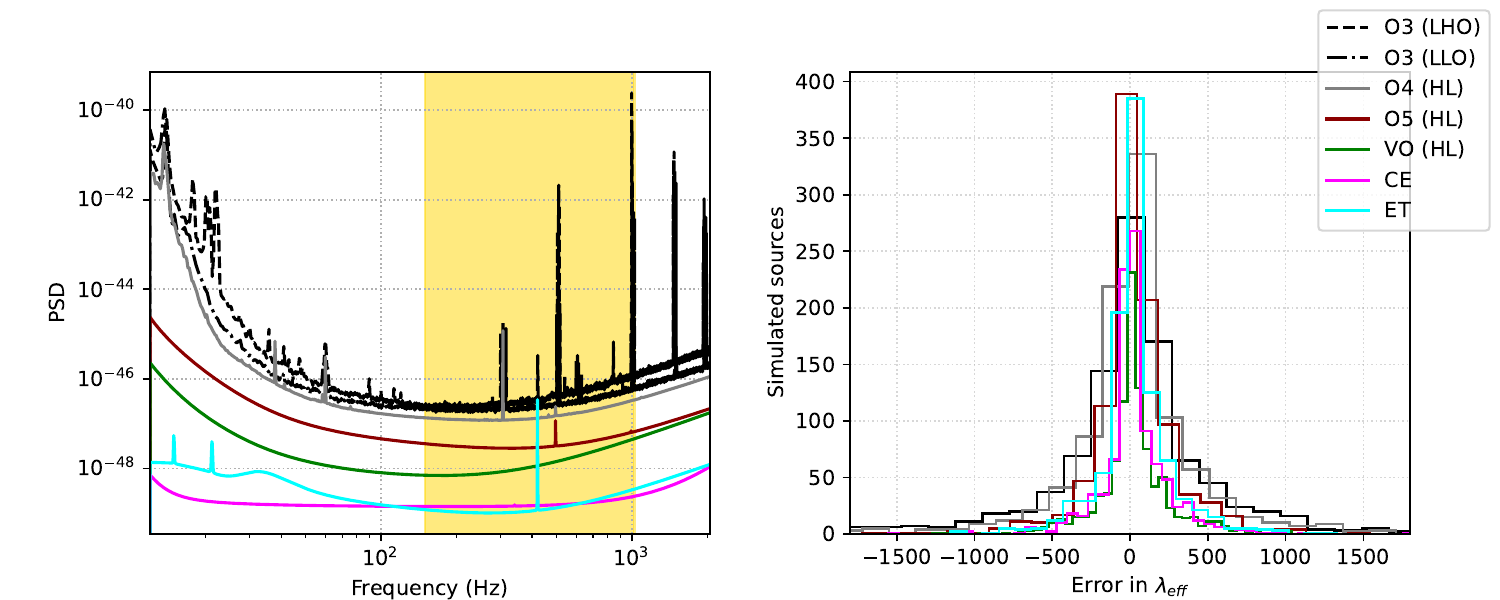}
\par\end{centering}
\caption{Errors in the point estimates for 1k simulated sources with different PSDs in the frequency band 150 Hz - 1024 Hz (in yellow). The method provides unbiased estimates throughout various observing scenarios but in general, provides more accurate estimates as the sensitivies in the relevant frequency band gradually improve.\vspace*{0.5cm}}
\label{fig:errors_det}
\end{figure*}
\subsection{Current observations}\label{sec:currentobs}
Here we are interested in the BNS signals from the Advanced LIGO and the Advanced Virgo observing runs, whose extrinsic parameters like the sky location, the luminosity distance and the inclination angles are approximately known from observations. The BNS events reported so far are \texttt{GW170817} and \texttt{GW190425}. The latter was observed only by the LIGO Livingston observatory and no electromagnetic counterpart could be observed. On the other hand, the former event was recored at both the LIGO Hanford and the LIGO Livingston observatories with a coincident network SNR $>$ 32 in the $\sim$ 23 Hz - 1024 Hz band, making it the loudest GW event till the end of O3. The HL coincident SNR of $\sim$ 16 in the $\sim$ 150 Hz - 1024 Hz band apparently makes it a suitable candidate for our analysis. A sub-threshold signal was also observed in Virgo improving the sky localization of the source. The sky location and the distance information for the source are also approximately known from other astronomical (EM) observations~\citep{Soares-Santos_2017,Cantiello_2018}. We conduct the multi-swarm analyses as described earlier with the GW data $\sim$ 4 seconds close to the merger time of the event with the goal of getting insights on its tidal effects.

While we expect that the estimation of $M\textsubscript{chirp}$ and $\chi\textsubscript{eff}$ are more accurate in the full frequency range, this analyses still provides comparable estimates with the previously reported results. The component mass estimates are consistent with the $M\textsubscript{chirp}$-uncertainty contour, which we expect to be slightly broadened due to the smaller signal strength in the chosen analysis band, whereas the estimated $\chi\textsubscript{eff}$ is fairly consistent with a nominal value as shown in Fig. \ref{fig:GW170817_multi_swarm}. As demonstrated in the earlier section, $\lambda\textsubscript{eff}$ which is expected to be better estimated with a higher $f\textsubscript{lower}$ than that with a lower $f\textsubscript{lower}$, is significantly different from that of the previously reported results~\citep{PhysRevX.9.011001, dai2018parameter, Radice, PhysRevLett.121.091102}. However, we also find that our estimate is prone to variations of random noise origin and the inference could be dependent on the specific noise realization of the instruments at the time of recording of the signal as shown in the Fig.~\ref{fig:noise_inference}. Thus, at the given sensitivities of the current detectors, our inference could not establish the true estimate of $\lambda\textsubscript{eff}$ with certainty, primarily due to the effect of the noise realization~\footnote{Notably, in the full Bayesian setting, the peak of the posterior can miss the true values at the current or past sensitivities as discussed in~\citep{PhysRevD.89.103012}.}. Either multiple such observations are needed to reduce the effect of noise realizations or exceptionally loud events can possibly inform us about the true estimates in the future e.g. with a third generation (3G) detector network. Thus, we do not explore further implications on the EoS of neutron stars based on the above inference. However, we conclude that a harder EoS may not be strictly ruled out based on the few observed neutron star binaries. This may be in contradiction to the general preference for a softer EoS in the community. The possibility of larger $\lambda\textsubscript{eff}$ values provide stronger evidence for the nature of the components away from BHs from GW observations, which has been already confirmed through the EM observations in this case.
\subsection{Future observations}
In the next generation of ground-based GW detectors, the rate of detections is likely to increase depending on the astrophysical abundancy of the sources and the factors like the sensitivity and the uptime of the instruments. The total SNR of a substantial fraction of the detectable events could be significantly large, given the number of low-frequency inspiral cycles that the plausible $f\textsubscript{lower}$-s would contain. Roughly, a signal from $1 M_{\odot} - 1 M_{\odot}$ would last for more than 10,000 seconds in a detector's sensitive frequency band starting at, say 5 Hz. Data analysis, including the standard Bayesian parameter estimation, of such long signals is a general challenge for GW astronomers. Since the approach presented here uses a small duration of the data near the merger, the analysis is less prone to be affected by glitches and other short transient signals that may occur in the middle of a long inspiral. The requirement of small duration of data in the analysis also avoids the need to consider the rotation effects of the earth during the full inspiral. We expect the spin-tidal measurement correlation to be more pronounced in the full frequency band analyses with the improved low frequency sensitivities. Thus the method presented here can be promising, however, to include the related effects at further higher frequencies, a higher sampling rate would be considered (e.g. to increase $f\textsubscript{upper}$ from 1024 Hz to 2048 Hz, we just need to increase the sampling rate from 2048 Hz to 4096 Hz~\footnote{which would roughly be just twice as costly as presented here; the overall cost of the analyses would still remain low since a small duration of data is used}). The point estimates from this approach are expected to become more accurate as the overall sensitivities improve as shown in Fig.~\ref{fig:errors_det}. Note that the $M\textsubscript{chirp}$ information for low mass binaries is a fairly accurate byproduct of the detection algorithms. The technique proposed here could be used to aid source classification in low-latencies and possibly lead to data products that provide valuable insights on the EM properties of the sources.

\begin{figure*}[t]
\begin{centering}
\includegraphics[scale=0.56]{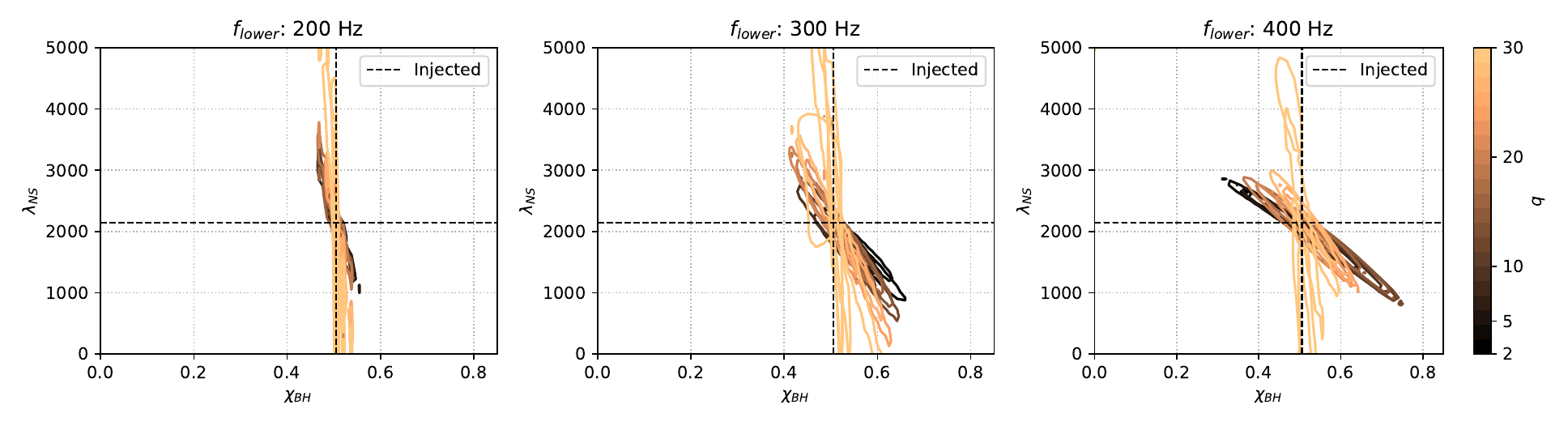}
\par\end{centering}
\caption{Simulation of NSBH systems differing in the BH mass with the masses of the NS is fixed at $1.4 M_{\odot}$ for varying $f\textsubscript{lower}$'s. The contours represent top 97\% values of the likelihood function computed in a similar fashion as in Fig. \ref{fig:lamvschi}. We use the \texttt{IMRPhenom\_NSBH} waveform model for generating the injection and the templates.\vspace*{0.5cm}}
\label{fig:lamvschi_NSBH}
\end{figure*}
\section{Conclusions}
This study shows that ignoring large generic aligned-spins in the presence/absence of tides in the detection stage can lead to sub-optimal recovery of neutron star signals. However, the measurement of the effective spin parameter of the neutron star components can affect the measurement of the effective tidal parameter in the full frequency bandwith of the detectors. We have discussed a novel technique to extract the tidal information primarily from the late-inspiral part of neutron star mergers signals observable with the ground-based detectors. Though a restricted frequency band reduces the recovered SNR, we find the strategy effective in probing the tidal deformability effects in binary systems. Using particle swarm optimization (PSO), we have explored an empirical low frequency cutoff, $f\textsubscript{lower}$ for the signals that minimizes the errors in the point estimates of the effective tidal parameter, $\lambda\textsubscript{eff}$. We observed that a frequency band around 150 Hz - 1024 Hz efficiently provides more reliable estimates for a test population of simulated binary neutron stars than the full frequency band. We also found that this technique benifits from the numerical relativity (NR) informed waveform models which excel in modeling the near-coalescence regime.

While in this approach, the estimate is originally obtained as a point realization in the parameter space, it can be conducted repeatedly (given its low computational cost) and combined to obtain the uncertainty. Since this is originally a frequentist idea, a natural question is regarding what to expect from a corresponding Bayesian analysis in a $\sim$150-1024 Hz band. In general, one anticipates broadening of a posterior in a reduced frequency range due to a smaller SNR than that in the full frequency band. Thus, for loud sources at neglibible noise realization effects, it may be reasonable to expect that while the posterior distribution broadens, it still peaks at the true $\lambda\textsubscript{eff}$. It is unclear though how the spin-tide correlation manifests itself through Bayesian approaches. Hence, studies with the Bayesian methods in a reduced frequency band are also worthwhile.

We have studied the effect of detector noise realizations on a frequentist's inference of the effective tidal deformability. The inference obtained with the method described in this work are affected by the detectors' noise realizations at the current sensitivities and thus could not provide a robust estimate of $\lambda\textsubscript{eff}$ for \texttt{GW170817}. However, we conclude that for similar events in the future, the estimates are expected to become free from uncertainties due to the random noise effects in the upcoming observing scenarios. Besides, the new approach could be especially beneficial in these future observations when the duration of these signals would be longer (and hence, computationally challenging) with the improved/proposed low frequency sensitivities of the current and the upcoming ground-based detectors.

The study can be extended further to understand the correlation with/among other effects such as eccentricity and spin-precession. Further investigations with the above strategy are desirable for binaries containing sub-solar mass (SSM) objects or black holes (BH) accompanying the neutron stars. In the case of neutron star-black hole (NSBH) systems, we observe that the tidal deformability of the NS tends to depend on the mass of its BH counterpart, as shown in Fig. \ref{fig:lamvschi_NSBH}. It is believed that the tidal deformability of a BH is zero which can be used as a test of the BH-nature of a component. However, the BH can cause measurable tidal deformability to the accompanying NS under some circumstances. As demonstrated earlier in this work, the tidal component is almost uninforming at $f\textsubscript{lower} \sim 20$ Hz and starts constraining at a higher $f\textsubscript{lower}$. Nevertheless, we find that even at a higher $f\textsubscript{lower}$, the contours tend to be unconstrained for large mass ratios ($q$) systems, possibly indicating the absence of a measurable tidal deformability. Because a large mass BH would merge quickly with a typical NS, thus allowing almost no time to tidally deform and disrupt the NS. Generally, the tidal interactions can be expected when the BH mass is comparable to the NS mass, while also increasing the chances for an EM emission. We also encourage the readers to refer to an interesting discussion in the following review article~\citep{NeutronKaterina}.
\section{Acknowledgments}
This research has made use of data or software obtained from the Gravitational Wave Open Science Center (gwosc.org), a service of LIGO Laboratory, the LIGO Scientific Collaboration, the Virgo Collaboration, and KAGRA. LIGO Laboratory and Advanced LIGO are funded by the United States National Science Foundation (NSF) as well as the Science and Technology Facilities Council (STFC) of the United Kingdom, the Max-Planck-Society (MPS), and the State of Niedersachsen/Germany for support of the construction of Advanced LIGO and construction and operation of the GEO600 detector. Additional support for Advanced LIGO was provided by the Australian Research Council. Virgo is funded, through the European Gravitational Observatory (EGO), by the French Centre National de Recherche Scientifique (CNRS), the Italian Istituto Nazionale di Fisica Nucleare (INFN) and the Dutch Nikhef, with contributions by institutions from Belgium, Germany, Greece, Hungary, Ireland, Japan, Monaco, Poland, Portugal, Spain. KAGRA is supported by Ministry of Education, Culture, Sports, Science and Technology (MEXT), Japan Society for the Promotion of Science (JSPS) in Japan; National Research Foundation (NRF) and Ministry of Science and ICT (MSIT) in Korea; Academia Sinica (AS) and National Science and Technology Council (NSTC) in Taiwan.

This material is based upon work supported by NSF's LIGO Laboratory which is a major facility fully funded by the National Science Foundation. The article has a LIGO Document number LIGO-P2400130. We have primarily used the PyCBC software in conducting various analyses in this work. All the waveform models used in this work are publicly available and are implemented in \texttt{LALSuite}~\citep{lalsuite}. We acknowledge the use of the Sarathi computing cluster at IUCAA for computational/numerical work. A significant part of the numerical work was also carried out at the Center of Excellence in Space Sciences, India (CESSI). CESSI is a multi-institutional Center of Excellence hosted by the Indian Institute of Science Education and Research (IISER) Kolkata and has been established through funding from the Ministry of Education, Government of India. SP wants to thank the financial support from the Council for Scientific and Industrial Research (CSIR), India through File No:09/921(0272)/2019-EMR-I. We sincerly appreciate the valuable inputs received from the anonymous referee.
\bibliography{tid_pso}
\bibliographystyle{aasjournal}
\end{document}